\documentclass[journal=jacsat,manuscript=article,onecolumn,showpacs,preprintnumbers,amsmath,amssymb,amsfonts]{achemso}

\everymath{\displaystyle}

\usepackage{longtable}

\usepackage[version=3]{mhchem} 

\def\be{\begin{equation}}
\def\ee{\end{equation}}
\def\bea{\begin{eqnarray}}
\def\eea{\end{eqnarray}}

\author{Anirban Polley$^1$}
\author{Satyajit Mayor$^2$}
\author{Madan Rao$^{1,2}$}
\email{madan@rri.res.in, madan@ncbs.res.in}
\affiliation[University]
{$^1$Raman Research Institute, C.V. Raman Avenue, Bangalore 560080,
India\\
$^2$National Centre for Biological Sciences (TIFR), Bellary Road,
Bangalore 560065, India}

\title[\texttt{achemso} demonstration]
{Bilayer registry in a multicomponent asymmetric membrane : dependence on lipid composition and chain length}

\begin{document}

%
%
%

\begin{abstract}
A question of considerable interest to cell membrane biology is whether phase segregated domains across an asymmetric bilayer are strongly correlated with each other and whether phase segregation in one leaflet 
can induce segregation in the other. We answer both these questions in the affirmative, 
using an atomistic molecular dynamics simulation to study the equilibrium statistical properties of a 3-component {\em asymmetric} lipid bilayer 
comprising  an unsaturated POPC (palmitoyl-oleoyl-phosphatidyl-choline), a saturated SM (sphingomyelin) and cholesterol with different composition ratios. Our simulations are done by fixing the
composition of the upper leaflet to be at the coexistence of the liquid ordered ($l_o$) - liquid disordered ($l_d$) phases, while the composition of the lower leaflet is varied  from the phase coexistence 
regime to the mixed $l_d$ phase, across a first-order phase boundary.
In the regime of  phase coexistence in each leaflet, we find strong  transbilayer correlations of the $l_o$ domains across the two leaflets, resulting
in  {\it bilayer registry}. This transbilayer correlation depends sensitively upon the chain length of the participating lipids and possibly other features of lipid chemistry, such as degree of saturation. We find that the $l_o$ 
domains in the upper leaflet can {\em induce} phase segregation in the lower leaflet, when the latter is nominally in the mixed ($l_d$) phase.
\end{abstract}
\maketitle

\section{Introduction} 

Cell membranes are composed of many different lipid species and exhibit both lateral heterogeneity \cite{simons1,simons_science2010,simons_nat_rev_2000} and bilayer asymmetry in their lipid composition \cite{TrafficRaoMayor}. While there have been many in-vitro studies of lateral phase segregation in multicomponent  giant unilamellar vesicles (GUVs)  \cite{keller,keller03,webb} and suspended membranes  \cite{keller,keller03}, it is only recently that attention has turned to membranes with asymmetric bilayers \cite{wan,SarahKeller_PNAS2009}. One of the issues  highlighted in these experiments and relevant to 
cell membrane biology, is the extent of correlation or registry of  phase segregated domains in the two leaflets of the
bilayer. This has inspired theoretical  \cite{schick,schick_BPJ2011} and coarse-grained computer simulation  \cite{marrink_PNAS2008}
studies of {\it bilayer registry} of domains in asymmetric bilayers. A well studied multicomponent model system is the 3-component lipid mixture comprising an unsaturated lipid (POPC), a saturated
lipid (PSM) and cholesterol (Chol) which exhibits liquid-ordered ($l_o$) - liquid-disordered ($l_d$) phase coexistence. Since the extent of bilayer registry is likely to be 
sensitive to lipid chemistry, in this paper we study the transbilayer coupling and extent of bilayer registry of the phase domains across the membrane,
using an atomistic molecular dynamics (MD) simulation of an asymmetric lipid bilayer membrane comprising POPC/PSM/Chol.

%

%

The registry of lipids across the bilayer suggest a mechanism by which outer leaflet lipids may couple with inner leaflet lipids and vice versa. This is important in the construction of membrane domains, and more generally in transducing information across the bilayer by lipidic receptors such as GPI-anchored proteins (GPI-APs) \cite{GPI-receptors} or glycolipids \cite{glycolipids} and other lipid species. 
Our motivation for this work also comes from  a series of experiments that study the spatial organization and dynamics of GPI-APs, on the surface of 
 living cells.
A variety of experimental strategies such as  Fluorescence Resonance Energy Transfer (FRET) \cite{TrafficRaoMayor,raftreviews,sharma,debanjan,kripa}, near-field microscopy (NSOM)
and electron microscopy, have revealed that both the organization and dynamics of the outer-leaflet GPI-APs are regulated by cholesterol, sphingolipids and {\it cortical actin and myosin} at the inner leaflet. The 
question is how do the outer-leaflet GPI-APs couple to cortical actin that abuts the inner leaflet of the cell membrane \cite{debanjan,kripa}. Since the interaction across the bilayer must be indirect,
are there other lipids, such as cholesterol and sphingolipids, that are involved in this linkage ? Do specific inner leaflet lipids that interact with actin,
 participate in this transbilayer coupling 
\cite{kripa} ? This naturally brings up the issue of bilayer registry in the cell membrane and its dependence on the specificity of lipids and its chemistry. We have been addressing these
important issues using both experiments on live cells and atomistic molecular dynamics simulations on multicomponent model membranes composed of so-called `raft-components'  \cite{simons1}.

The article is organized as follows : we first describe the details of the atomistic molecular dynamics (MD) simulation of the 3-component bilayer. Next we present our results
on lateral compositional heterogeneity, extent of bilayer registry and mismatch area across the bilayer, as a function of the 
concentration of the saturated lipid (PSM). We also study how changes in lipid chain length of SM
affect bilayer registry. We end with a short summary of the results and conclusions.

\section{Methods}

\noindent
{\it Model membrane}\,:\,We study the phase segregation and bilayer registry of a symmetric and asymmetric  3-component bilayer membrane embedded in an aqueous medium by
atomistic molecular dynamics simulations (MD) using {\it GROMACS}. 
We prepare the bilayer membrane at  $23^{\circ}$C at different relative concentrations of 
palmitoyl-oleoyl-phosphatidyl-choline (POPC), long chain palmitoyl-sphingomyelin (SM-16:0) (PSM) and cholesterol (Chol). 
All multicomponent bilayer membranes have $512$ lipids in each leaflet (with a total $1024$ lipids) and $32768$ water molecules (such that the ratio of water to lipid is $32:1$) 
so as to completely hydrate
the simulated lipid bilayer. 

For the symmetric bilayer, the 
relative concentration (in percentage, $x$) of PSM and Chol in  the upper and lower leaflet is 
varied from  $x \,(in\, \%)=33.3, 19.9, 12.5,10.0, 9.1, 7.1, 5.8, 2$ and $1\%$, with 
POPC contributing to the rest of the lipid content.

\begin{figure*}[h!t]
\includegraphics[width=16.0cm]{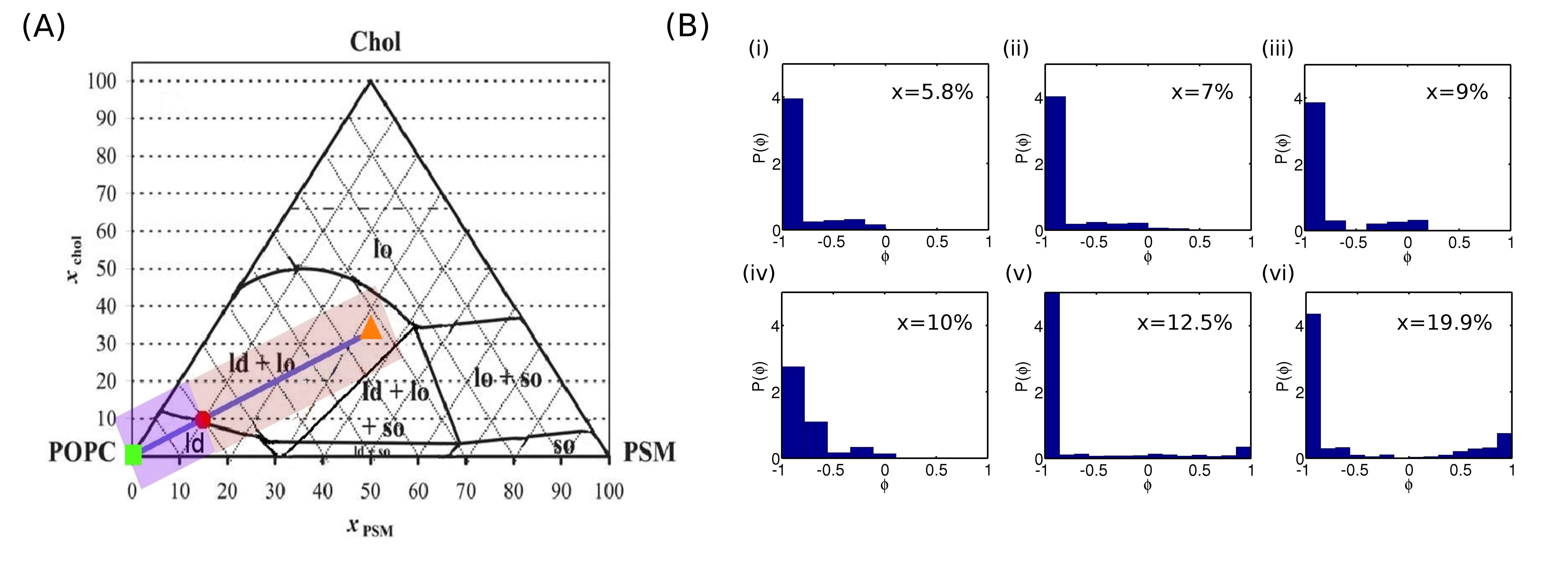}
\caption{(A) Ternary phase diagram of POPC, PSM and CHOL at $T= 23^{\circ}$C taken from Ref.\,[25]. Triangle (orange) represents the composition  $1:1:1$ which is deep in the phase coexistence region; Dot (red) is a point on the phase boundary $x_c = 10\% $. 
(B) We have verified this phase diagram by doing simulations at different compositions along the blue line in (A). Panel shows the probability distribution of the phase segregation order parameter $\phi$ (main text) for the symmetric bilayer at different values of $x$ : (i) - (iii)  shows that for $x<10$\%, the bilayer is in the $l_d$ phase, while (v) - (vi) shows that for $x > 10$\%, the membrane is at $l_o$-$l_d$ coexistence. The
phase transition is  clearly at (iv) $x_c=10$\%.
}
\label{Fig1}
\end{figure*}

\begin{figure*}[h!t]
\includegraphics[width=16.5cm]{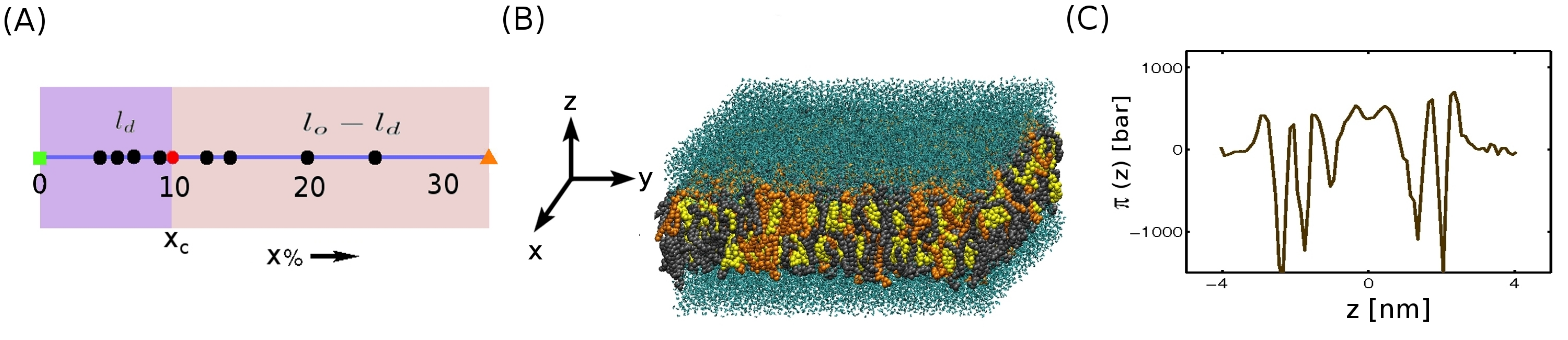}
\caption{(A) For the simulations of the asymmetric bilayer,  we hold the composition of the upper leaflet at $1:1:1$, while in the lower leaflet the composition of PSM and Chol is varied from $x=33.3\%$ (orange dot) to $0\%$ (green square), with POPC contributing to the rest. The composition where the simulations are carried out, denoted by black dots (see {\em Methods}), are indicated against the reference ternary phase diagram of the symmetric 
bilayer (blue line in Fig.\,1A). Triangle (orange) represents the composition  $1:1:1$ which is deep in the phase coexistence region; Dot (red) is a point on the phase boundary $x_c = 10\% $. 
(B) Snapshot of equilibrium configuration of the asymmetric bilayer when $x=25\%$, with POPC (gray), PSM (orange), Chol (yellow) and water (cyan). (C) Lateral pressure profile $\pi(z)$ for the same bilayer at equilibrium. 
}
\label{Fig2}
\end{figure*}

For the asymmetric bilayer,  the upper leaflet has 
POPC /PSM /Chol in the ratio $1:1:1$ (i.e., the relative concentration of each component is $33\%$). We vary the relative composition in the lower leaflet; denoting $x$ as the 
relative concentration (in percentage) of PSM and Chol in the lower leaflet, we run through the values $x \,(in\, \%)=33.3, 25.0, 19.9, 14.3, 12.5,$ $10.0, 9.1, 7.1, 5.9$ and $ 4.5 \%$, with 
POPC contributing to the rest of the lipid content.
With this choice of compositions and temperature, the upper leaflet is in the putative $l_o$-$l_d$ phase coexistence regime (see \ref{Fig1}A, for the ternary phase diagram at $23^{\circ}$C, taken from Ref.\,[17]), while in the 
lower leaflet the compositions straddle the phase boundary allowing us to go from the $l_o$-$l_d$ phase coexistence regime to the $l_d$ phase, \ref{Fig2}A.

To study the role of  lipid chemistry, we repeat the above set of simulations with PSM in the lower leaflet replaced by the short chain sphingomyelin, SM-14:0 (MSM). We  
vary  the concentration $x$  of MSM (Chol) across the range $x\, (in \, \%) =4.5, 5.9, 7.1, 9.1, 10.0, 12.5$ and  $14.35\%$. 
\\

\noindent
{\it Force fields}\,:\,The force field parameters for POPC, PSM and Chol are taken from the previous validated united-atom description \cite{anirban_jpcb12,Tieleman-POPC,mikko}. We construct the force field parameters for MSM (SM-14:0) from the parameters of PSM and POPC. We use the improved extended simple point charge (SPC/E) model to simulate water molecules, having an extra average polarization correction to the potential energy function.    
\\

\noindent
{\it Initial configurations}\,:\,We generate the initial configurations of the asymmetric multicomponent bilayer membrane using {\it PACKMOL} \cite{packmol}.
For all simulation runs, we choose two sets of initial conditions : (i) where the components in each leaflet are homogeneously mixed and (ii) where the ternary components are completely phase segregated
in $l_o$-$l_d$ domains \cite{anirban_jpcb12}.
\\

\noindent
{\it Choice of ensembles and equilibration}\,:\,The asymmetric bilayers are equilibrated for $50$\,ps in the NVT ensemble using a Langevin thermostat to avoid bad contacts arising from steric constraints and then for $160$\,ns in the NPT ensemble ($T = 296$\,K ($23^{\circ}$C), $P =1$\,atm). The simulations are carried out in the NPT ensemble for the first $20$\,ns using Berendsen thermostat and barostat, then for $20$\,ns using Nose-Hoover thermostat and the Parrinello-Rahman barostat to produce the correct ensemble. Rest of the simulations are performed in the NPT ensemble using Berendsen thermostat. We use a semi-isotropic pressure coupling with compressibility $4.5\times 10^{-5}$ bar$^{-1}$ for the simulations in the NPT ensemble. 

The long-range electrostatic interactions are incorporated by the reaction-field method with cut-off $r_c = 2$\,nm, while for the Lennard-Jones interactions we use a cut-off of $1$\,nm
\cite{anirban_jpcb12,mikko,patra2004}.

For each initial configuration, we run the simulations for $200$\,ns before computing the desired physical quantities. 
To ensure that the bilayer membrane is well equilibrated, we monitor the  area per lipid throughout the simulations ({\em Supplementary Figure S1-S2}). We calculate the lateral pressure profiles in the bilayer using Irving-Kirkwood contour and grid size $0.1$\,nm. We calculate the pairwise forces by rerunning the trajectory with cut-off  $2$\,nm for electrostatic interactions using LINCS algorithm to constrain the bond lengths \cite{Hess} and the SETTLE algorithm to keep the water molecules rigid \cite{SETTLE} so that integrator time step of $2$\,fs can be used. We generate pressure profiles from trajectories over $20$\,ns using SHAKE algorithm \cite{SHAKE} to constrain bond lengths.
\\

\noindent
{\it Computation of deuterium order parameter}\,:\,We calculate the spatial distribution of the deuterium order parameter $S$ from the selected carbon atoms ($C5-C7$) of each acyl chain (including SN1 and SN2 chains) of the PSM and POPC lipids \cite{anirban_jpcb12,mikko}. Here, $S$ is defined for every selected CH$_2$ group in the chains as, $S \equiv   \frac{1}{2} \langle 3 \cos^2 \theta -1 \rangle$ where $\theta$ is the angle between a CH-bond and the normal to the plane of the membrane (z-axis). This is then coarse-grained (binned) over a spatial scale of $0.5$\,nm for last $20$\,ns of the trajectory of the simulations. We use our previous estimation of the deuterium order parameter $S$ of the $l_o$-$l_d$ domains of the bilayer membrane \cite{anirban_jpcb12}, to declare a region to be liquid-ordered ($l_o$) when the value of $S \ge 0.35$. \\

\noindent
{\it Computation of mismatch area}\,:\, 
We calculate the coarse-grained spatial profile of the deuterium order parameter, $S$ in each leaflet using grid size $0.5\,nm$. We then use the above cutoff in $S$ to declare a region as liquid ordered. We compute the area and perimeter of the $l_o$ domains in each leaflet using the cluster algorithm available in Image Processing Toolbox, MATLAB 2009. This is used to calculate the overlap and mismatch area of the domains
across the bilayer (see, Section {\it Mismatch area and interfacial tension}).

\section{Results and Discussion}

We compute the local stress profile of the bilayer membrane from the virial, and use this to calculate the net surface tension, force and torque. We ensure that the prepared bilayer membrane is mechanically stable, with both the net force and torque balanced. In addition we ensure that the surface tension is zero to within numerical error.  The details of the mechanically stable symmetric bilayer have appeared in an earlier
publication \cite{anirban_jpcb12}. \ref{Fig1}A shows the phase diagram of the symmetric bilayer comprising POPC, PSM and Chol at $23^{\circ}$C taken from Ref.\,[25]. 
We have simulated the symmetric bilayer membrane composed of POPC, PSM and Chol with concentration, $x=1\%, 2\%, 5.8\%, 7.1\%, 9.1\%, 10\%, 12.5\%, 19.9\%$ and $33.3\%$ of the PSM (Chol).
We have plotted $P(\phi)$ with different $x$ for the symmetric bilayer where, $\phi$ is defined as,
$\phi=\frac{\rho_{PSM}-\rho_{POPC}}{\rho_{PSM}+\rho_{POPC}}$ (\ref{Fig1}B).

For  details of the mechanical stability of the asymmetric bilayer, see  {\em Supplementary Tables S1-S6}, where we record the 
net force, torque and surface tension at each composition of the asymmetric bilayer in tabular form. Here, we show a snapshot of the ternary asymmetric bilayer membrane composed of POPC, PSM and Chol  and its lateral pressure profile $\pi(z)$, \ref{Fig2}\,B and C, respectively (profiles at other concentrations are displayed in {\em Supplementary Figure S3-S4}).


We perform simulations on our model asymmetric bilayer at varying concentrations $x$ of PSM and Chol in the lower leaflet, whilst maintaining  the upper leaflet at  a composition $1:1:1$,
which is deep in the $l_o$-$l_d$ phase coexistence region. The simulations done at various values of $x$ along the line shown in \ref{Fig2}A, traverses across the phase boundary at $x_c=10\%$ into the $l_d$ phase.



We perform a similar study when the lower leaflet PSM is replaced by the short chain sphingomyelin, MSM.



\begin{figure*}[h!t]
\includegraphics[width=14.0cm]{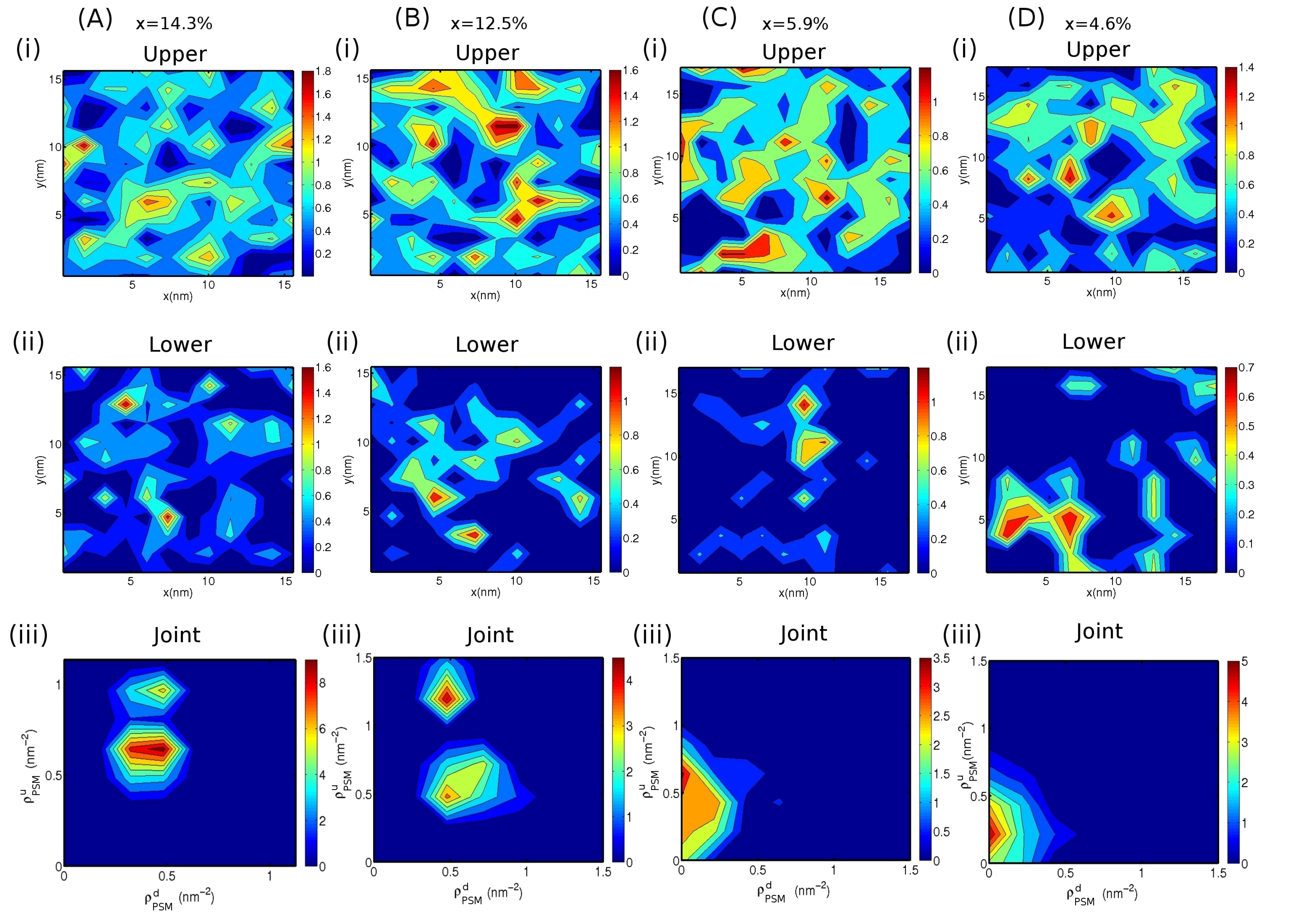}
\caption{ 
Spatial profile of the coarse-grained number density of PSM (color bar) in the upper leaflet (top panel) at different compositions of PSM in the lower leaflet (as indicated at the top of the panels) 
The middle panel shows the corresponding profile of the number density of PSM (color bar) in the lower leaflet. The bottom panel shows the extent of correlation between the PSM-rich domains across the bilayer, as measured by the
  joint probability distribution, JPD (color bar, see text). For $x>x_c=10\%$, the JPD show strong transbilayer correlations, while for $x \ll x_c$, the correlations are poor.
}
\label{Fig3}
\end{figure*}

\subsection{Lateral compositional heterogeneity}
The coarse-grained spatial profile of the lipid number density is calculated with a grid size $1.3$\,nm. As stated in {\it Methods}, the composition in the upper leaflet is fixed at $1:1:1$, while the
composition of PSM/Chol in the lower leaflet is varied from $33\%$ to $4.5\%$. The top and middle panels in \ref{Fig3} show the spatial profile of the number density of PSM in the upper and lower leaflets, respectively,
 at 4 representative compositions on either side of the phase boundary, $x_c=10\%$. The lower panel, described in the next section, shows the joint correlation of the PSM rich domains across the bilayer.

\begin{figure*}[h!t]
\includegraphics[width=14.0cm]{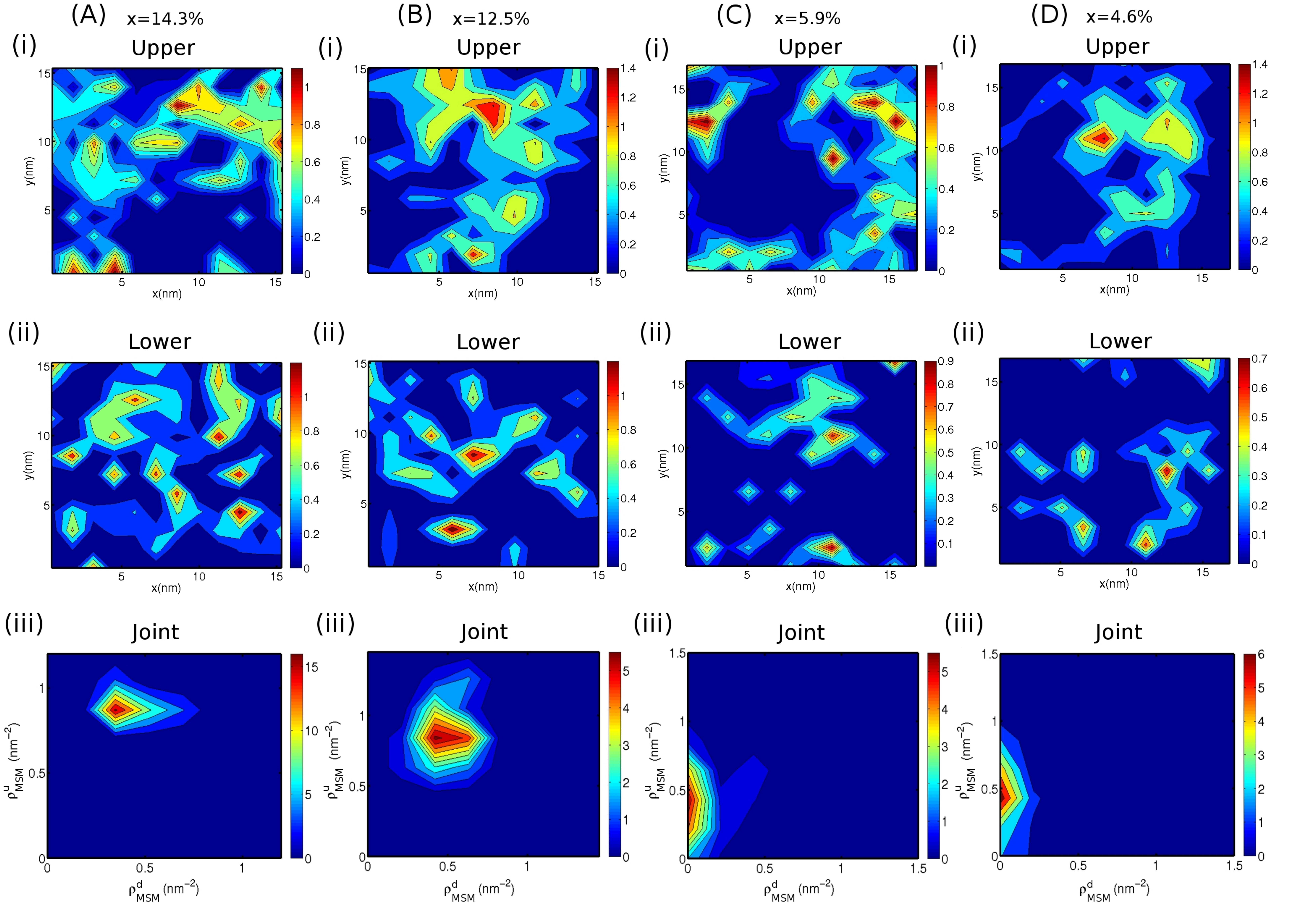}
\caption{
Spatial profile of the coarse-grained number density of PSM (color bar) in the upper leaflet (top panel) at different compositions of MSM in the lower leaflet (as indicated at the top of the panels) 
The middle panel shows the corresponding profile of the number density of MSM (color bar) in the lower leaflet. The bottom panel shows the extent of correlation between the PSM-rich domain in the upper leaflet and the MSM-rich domain in the lower leaflet, as measured by the joint probability distribution, JPD (color bar, see text). As in \ref{Fig3}, for $x>x_c=10\%$, the JPD show strong transbilayer correlations, while for $x \ll x_c$, the correlations are poor.
}
\label{Fig4}
\end{figure*}

\ref{Fig4} shows a similar study done when PSM in the lower leaflet is replaced by short chain MSM.

%
%

\subsection{Domain registry across bilayer}
We have studied the extent of registry of $l_o$-$l_d$ domains across the bilayer of an asymmetric multicomponent membrane as a function of varying composition and lipid chemistry.  We measure the extent of transbilayer registry by computing the joint probability distribution (JPD) of the 
coarse-grained number density of PSM in the upper and lower leaflets at the same coarse-grained spatial location $(x,y)$. The lower panel of \ref{Fig3} shows the JPD at different values of the concentration $x$ of
PSM in the lower leaflet.



In the lower panel in \ref{Fig3}, the JPD shows a distinct peak along the diagonal when $x>x_c=10\%$,
which is clear evidence of {\it bilayer registry} of $l_o$-domains. The off-diagonal peak in the JPD is merely an indication of the relative abundance of PSM in the upper leaflet.
On the other hand, for $x\ll x_c$, this diagonal peak in the JPD is absent, indicating lack of bilayer registry.

A similar conclusion regarding the bilayer registry can be drawn when the lower leaflet PSM is replaced by the short chain MSM (\ref{Fig4}).

These observations suggest that the configurations of the two leaflets mutually influence each other. As stated in the {\it Abstract}, we can ask whether the segregation of lipids in the upper leaflet can induce a phase segregation in
lower leaflet, i.e., can the composition in the upper leaflet act as a local ``field'' for the composition in the lower leaflet. To study this, we define a  `transbilayer order-parameter' 
from the normalized transbilayer correlation ($r$ denotes the 2d coordinate $(x,y)$),

\be 
C(\rho^{u}_{PSM}(r),\rho^{d}_{PSM}(r))=\frac{\langle \rho^{u}_{PSM}(r) \rho^{d}_{PSM}(r) \rangle - \langle \rho^{u}_{PSM}(r) \rangle \langle \rho^{d}_{PSM}(r) \rangle}{\sqrt{\langle \rho^{u}_{PSM}(r)^2 \rangle - \langle \rho^{u}_{PSM}(r) \rangle^2} \sqrt{\langle \rho^{d}_{PSM}(r)^2 \rangle - \langle \rho^{d}_{PSM}(r) \rangle^2}} 
\ee
averaged over space (denoted by $C_{ud}$) and compute this as a function of the relative concentration $x$ of PSM/Chol. {\em Supplementary Figure S5} shows the transbilayer order-parameter 
$C_{ud}$ as a function of $x$ for a symmetric bilayer. $C_{ud}$ jumps from a high value in the $l_o-l_d$ phase coexistence region to a low value in the $l_d$ phase.
The  jump in $C_{ud}$ coincides with the phase boundary $x_c=10$\% (\ref{Fig1}).

\begin{figure}[h!t]
\includegraphics[width=14cm]{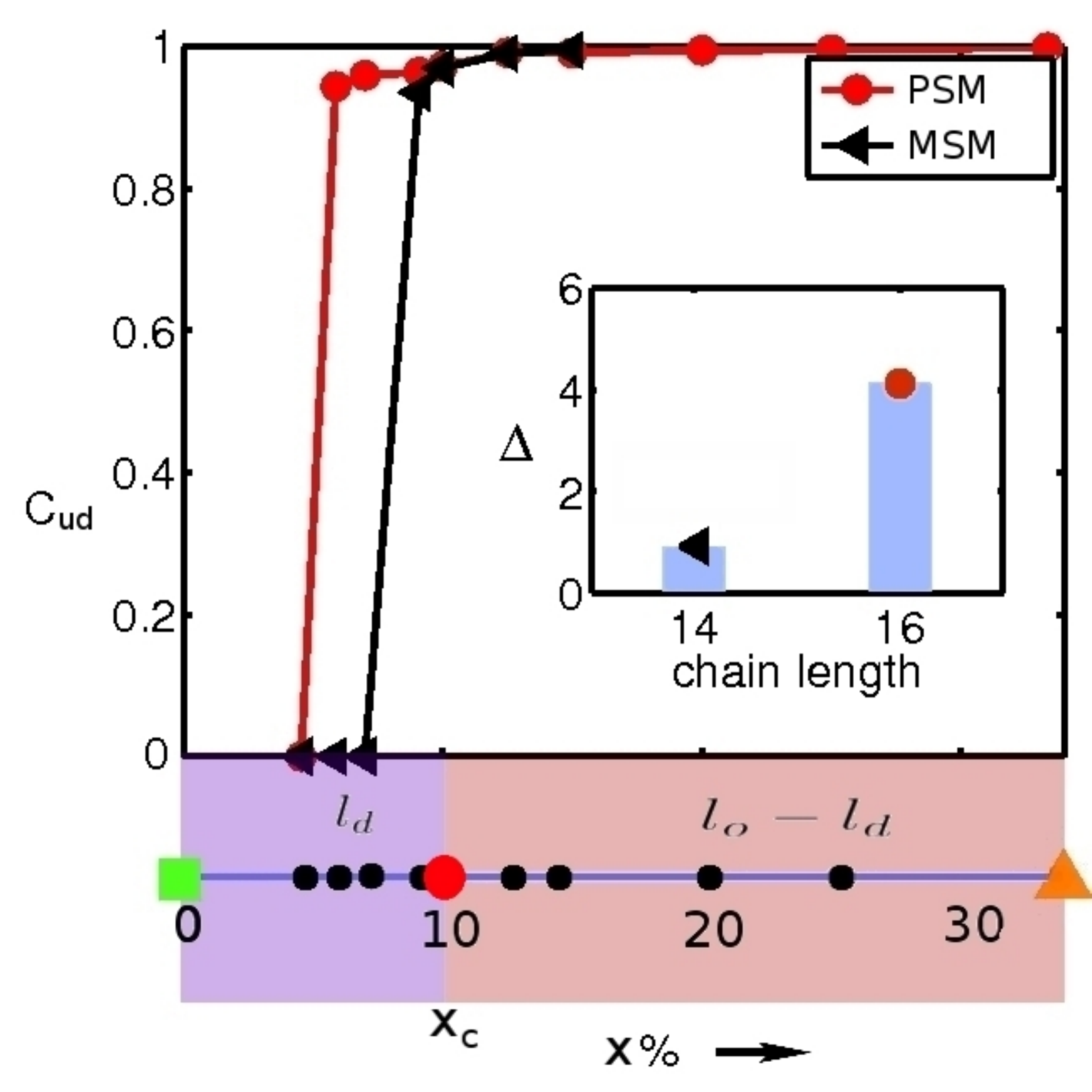}
\caption{Transbilayer order parameter $C_{ud}$ defined from the transbilayer correlation $C(\rho^{u}(r),\rho^{d}(r))$ (see text) between the density of upper leaflet PSM and lower leaflet PSM (red circle) or MSM (black triangle) versus $x$, the concentration of PSM or MSM in the
lower leaflet. Color panel below drawn for reference, denotes the values of $x$ at which $C_{ud}$ has been evaluated. The value of $C_{ud}$ is zero for small $x$ and jumps sharply at $x_c^{PSM/MSM} < x_c=10\%$ (red dot in color panel), indicating a first-order phase transition. The phase transition point for the long chained PSM, $x^{PSM}_c=5.88\%$ is smaller than that of the short chained MSM $x^{MSM}_c=9.09\%$. Inset shows the shift in the transition $\Delta$ (see text) as a function of lipid chain length.
}
\label{Fig5}
\end{figure}

For the asymmetric bilayer, we compute the transbilayer order parameter as a function of $x$, the concentration of PSM (or MSM) in the lower leaflet (\ref{Fig5}). 
The transbilayer order parameter $C_{ud}$  is very nearly zero for  $x\ll x_c$ and rises sharply to $\sim\,1$ at $x=x_c^{PSM} < x_c$, showing the influence of the upper leaflet on the phase segregation of the lower. This transbilayer influence is stronger for the long chain PSM than for the short chain MSM, as seen
by the fact that $x_c^{MSM} = 9.09\% > x_c^{PSM}=5.88\%$.


 There is thus  a shift in the phase boundary from its value of $x_c=10\%$ for the ternary symmetric bilayer of POPC-PSM-Chol. 
This shift is plotted as $\Delta=\vert x_c^{{PSM}/{MSM}}-x_c \vert$ for both the long chain PSM ($\Delta^{PSM}=4.12\%$) and short chain MSM ($\Delta^{MSM}=0.91\%$) in the lower leaflet (inset \ref{Fig5}). 


The above phenomenology can be understood within a mean-field theory of phase transitions \cite{Chaikin-Lubensky}, with a Helmholtz free-energy functional written in terms of $\phi_u$ and $\phi_d$, where $\phi_u$ ($\phi_d$) is the relative concentration of
the $l_o$ and $l_d$ species in the upper (lower) leaflet. The form of the free-energy functional for the asymmetric bilayer can be written as,
\bea
F[\phi_u, \phi_d] & = &  \frac{1}{2} \int d^2x \, \left[ f_u(\phi_u) + f_d(\phi_d) + f_{ud}(\phi_u, \phi_d)\right]   \nonumber \\
f_u &   =   & C_u (\nabla \phi_u)^2 - r_u \phi_u^2 + u \phi_u^3 + \textrm{v} \phi_u^4 \nonumber \\
f_d & = & C_d (\nabla \phi_d)^2 + r_d \phi_d^2 \nonumber \\
 f_{ud} & = & - A \phi_u \phi_d + B \phi_u \phi_d^2 + D \phi_u^2 \phi_d
 \label{free}
\eea
where $r_u \sim (x-x_c^u), r_d \sim (x-x_c^d)>0$ reflects the fact that the upper leaflet is in the $l_o$-$l_d$ phase coexistence regime, $\langle \phi_u\rangle \neq 0$, and the isolated lower leaflet is in the $l_d$ phase,  $\langle \phi_d\rangle = 0$. The coefficient $A>0$ to account for the fact that the local transbilayer coupling is attractive.

We first minimize $F$ with respect to $\phi_u$ : setting $\delta F/\delta \phi_u = 0$, and keeping terms to linear order, we get 
\be
C_u \nabla^2 \phi_u + r_u \phi_u = A \phi_d \,,
\label{EL}
\ee
whose Fourier transform, lends itself to a  useful interpretation, 
\be
\phi_u({\bf q}) = \frac{A \phi_d({\bf q})}{-C_u q^2 + r_u}\,,
\ee
namely a spatially varying $\phi_u$ can induce a spatially varying $\phi_d$. Nonlinearities in the free-energy that we have neglected, reinforce this 
and will lead to bilayer registry. Plugging this expression back into Eq.\,(3), we obtain an effective free-energy functional in terms of $\phi_d$ alone, which shows that the
coefficient of the quadratic term gets reduced by $r_d \rightarrow r_d - A^2/r_u$, which for large enough $A$ can become negative. This shows that the segregation in the
upper leaflet can induce a segregation in the lower, by shifting the phase transition point. This mean field analysis is entirely consistent with our MD simulations.


\subsection{Mismatch area and interfacial tension}
\begin{figure*}[h!t]
\includegraphics[width=14.0cm]{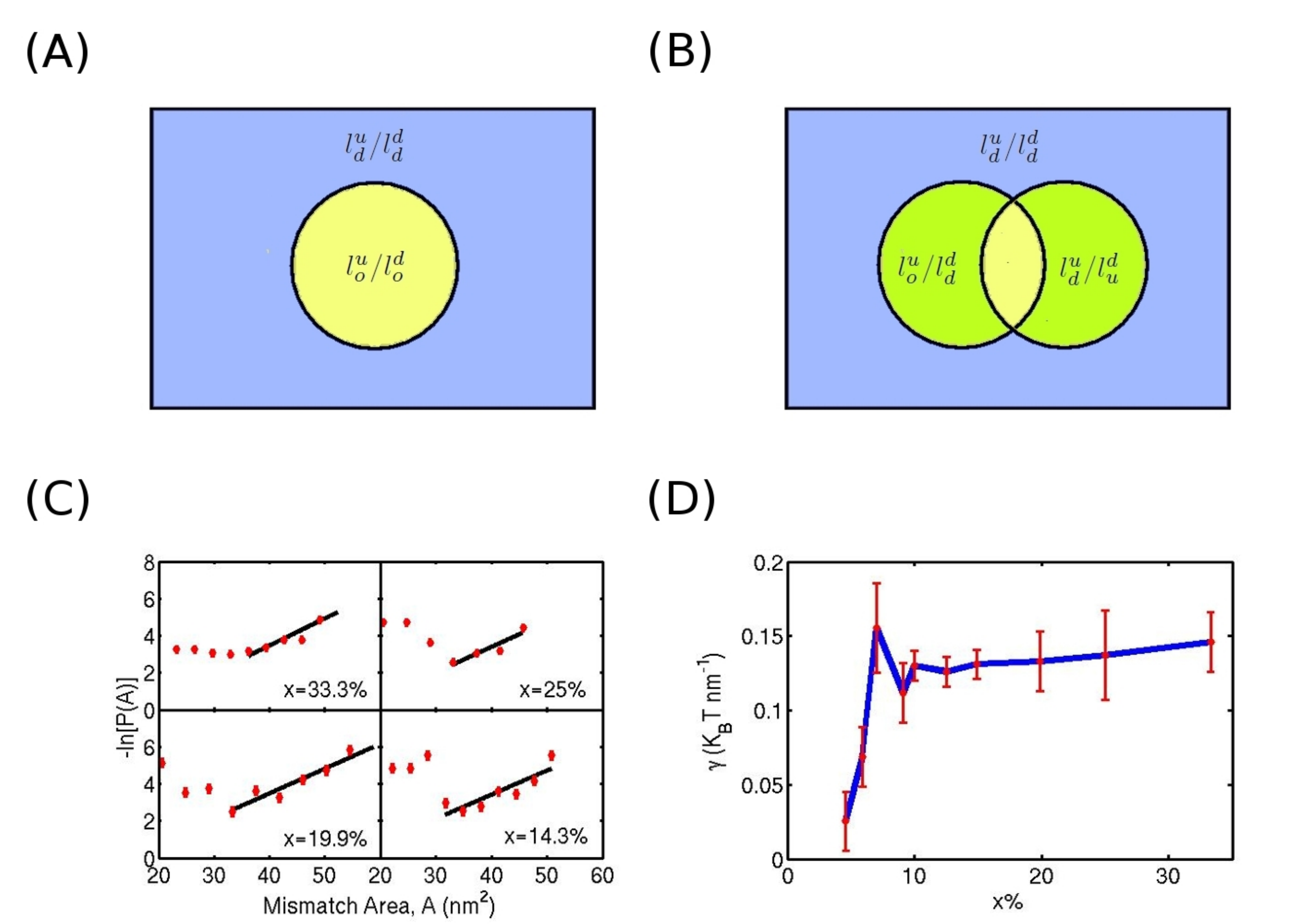}
\caption{ Schematic showing (A) domains in complete registry or overlap across the bilayer and (B) domains in partial overlap with a defined mismatch area (see text).
(C) Semi-log plot of the probability distribution of the mismatch area, $-\ln P(A)$ vs. $A$ (red dots), at different values of $x$, the concentration of PSM in the lower leaflet (indicated in the panel). The straight lines in the high $A$ regime are fits to the Boltzmann form (see text), from which we extract the value of the tension $\gamma$. Error bars are indicated.
 (D) Tension $\gamma$ as a function of $x$ shows a sharp drop to zero at $x\simeq x_c^{PSM}$.
}
\label{Fig6}
\end{figure*}

When there is  perfect bilayer registry, the area of the $l_o$ domain in the upper leaflet will completely overlap with the area in the lower leaflet (\ref{Fig6}A).
Any mismatch in the overlap area will cost energy proportional to the mismatch area $A$,
defined as $A=A^{u}_{l_o}+A^{d}_{l_o}-2 A^{o}$, where $A^{u}_{l_o}$ and $A^{d}_{l_o}$ are the areas of the $l_o$-domains in the upper and lower leaflets and $A^{o}$ is the overlap area between the $l_o$-domains in the upper and lower leaflets (\ref{Fig6}B). The proportionality constant is a tension $\gamma$ or a mismatch free energy per unit area, and is a measure of the domain overlap, a larger value of $\gamma$ implies a more complete overlap.  This tension $\gamma$ acts as a driving force for inter-leaflet registration of the phase domains across the bilayer. 
In principle, the value of the tension $\gamma$ is affected by short wavelength curvature and protrusion fluctuations, which we have ignored in our computation of the area - this will typically go to reduce the value of $\gamma$.



The  linear dependence of the energy on the mismatch area $A$ holds as long as the mean size of the mismatch region $\langle R\rangle$ is larger than its root mean square fluctuation  $w=\sqrt{\langle \delta R^2\rangle}$,
where $\delta R=R(\theta)-\langle R\rangle$, and $R(\theta)$ is the distance from the domain centre to the domain boundary at the angular position $\theta$.
There are strong corrections to this leading behaviour, of order $(w/\langle R\rangle)^2$, when the domains are small or ramified. Given that the lateral dimension of the model membrane is $15.6$\,nm, this is likely  the
case in our atomistic MD simulations. To check this, we have 
plotted the perimeter per area of the mismatch region ({\em Supplementary Figure S6}) versus area, at different values of $x$, the relative
concentration of PSM in the lower leaflet - this shows that the mean domain shapes deviate from 
circularity, especially for small values of $x$.


With this caveat, we have estimated the domain interfacial tension $\gamma$ by computing the probability distribution of the mismatch area $A$ of the $l_o$-domains between the two leaflets of the bilayer, and equating it to the
Boltzmann form, $P(A) \propto \exp({-\gamma A})$, where $\gamma$ is measured in units of $k_BT$. In {\em Supplementary Figure S7}, the plot of the probability distribution of $A$ at various values of $x$, shows a distinct peak at the
most probable value of $A$; in a semi-log plot \ref{Fig6}C, we fit the distribution to the Boltzmann form to extract the value of the tension $\gamma$. These values, at $x$ well within the coexistence region, for instance 
$\gamma=0.146\pm0.02$\,k$_B$T/nm$^{2}$ at $x=33\%$, are
consistent  with those obtained from other coarse-grained simulations \cite{Marrink_Risselada,Schick_Putzel}. Given the systematic errors in such a computation and the caveats mentioned above, we should regard this computed value of $\gamma$ with
some caution. Notwithstanding, the qualitative trend showing $\gamma$ decrease with $x$, with a sharp drop to zero at  $x \simeq x_c^{PSM}$ (\ref{Fig6}D), is reassuring.

  
\section{Conclusion}

We have analyzed the equilibrium properties of a ternary component, asymmetric bilayer membrane using atomistic molecular dynamics study. Our central goal was to study the conditions under which bilayer registry takes place in an asymmetric, multicomponent membrane. To summarize, our main results are: (i) $l_o$ phase domains formed in the two leaflets are registered across the bilayer membrane, (ii) phase segregation in upper leaflet can induce segregation in the lower, thus the composition on the upper leaflet acts as a ``field'' which couples linearly to the composition in the lower leaflet and (iii) the strength of the transbilayer coupling and the extent of bilayer registry depends sensitively on the lipid chain length and is greater for longer chain lipids.

The registry of the phase domains across the two leaflets of the bilayer membrane has an important implication to the sorting and signaling in live cell membrane. 
The cell membrane is inherently asymmetric with both lateral and transverse lipid heterogeneity. 
Recent experiments on live cells, using Fluorescence Resonance Energy Transfer (FRET) \cite{sharma,debanjan,kripa} show that outer leaflet GPI-APs organized as monomers and cholesterol-sensitive nanoclusters are regulated by the active dynamics of cortical actin (CA) and myosin. The present work forms the basis for further investigation of the transbilayer interaction between lateral heterogeneities of the outer leaflet GPI-anchored proteins, PSM and cholesterol with saturated, long chain lipids at the inner leaflet whose organization depends on the actin and actin remodeling proteins.
\\

\section*{Acknowledgements}
SM is a JC Bose Fellow (DST, Govt of India) and acknowledges support from an HFSP grant. This work was partially supported by a grant from Simons Foundation.


\end{document}